\documentclass{mpe_report}

\usepackage{psfig,graphicx,epsfig}
\usepackage{color}
\usepackage{amsmath,amssymb,epic,eepic,array}

\begin{document}

\pagenumbering{arabic}
\setcounter{page}{219}

 \renewcommand{\FirstPageOfPaper }{219}\renewcommand{\LastPageOfPaper }{222}
\title{Concepts for astronomical data accessibility and analysis
via relational database}
\author{L. Nicastro\inst{1} \and G. Calderone\inst{2}}  
\institute{Istituto Nazionale di Astrofisica, IASF,
 Via P. Gobetti 101, 40129 Bologna, Italy
\and Istituto Nazionale di Astrofisica, IASF,
 Via U. La Malfa 153, 90146 Palermo, Italy}
\maketitle

\begin{abstract}
Relational databases (DBs) are ideal tools to manage bulky and structured data
archives.
In particular for Astronomy they can be used to fulfill all the requirements
of a complex project, i.e. the management of: documents, software (s/w)
packages and logs, observation schedules, object catalogues, quick-look,
simulated, raw and processed data, etc.
All the information gathered in a relational DB is easily and
simultaneously accessible either from an interactive tool or a batch program.
The user does not need to deal with traditional files I/O or editing, but
has only to build the appropriate (SQL) query which will return the desired
information/data, eventually producing the aforementioned files or even
plots, tables, etc. in a variety of formats.
What is then important for a generic user is to have the tools to
easily and quickly develop, in any desired programming language,
the custom s/w which can
import/export the information into/from the DB. An example could be a
Web interface which presents the available data and allows the user
to select/retrieve (or even process) the data subset of interest.
In the last years we have been implementing a package called \textsf{MCS}
(see dedicated paper in this proceedings) which allows
users to interact with MySQL based DBs through any programming
language. MCS has a multi-thread (socket) architecture which means that several
clients can submit queries to a server which in turn manages the communication
with the MySQL server and other MCS servers.
Here we'll focus on a the real-world case of the robotic IR--optical telescope
REM (placed at La Silla, Chile) which performs real time images acquisition,
processing and archiving by using some of the MCS capabilities.
%
Interested people can visit \textsf{ross.iasfbo.inaf.it} to have a hint of
the potential of DB-based data management.
\end{abstract}

\section{Introduction}
Nowadays medium-large size astronomical projects have to face the management
 of a large amount of information and data. Typically dedicated data centres
 manage the collection of raw and pre-processed data and consequently make
 them accessible to the (authorized) users. Access is performed either via
 (s)ftp or http(s) (Web) and typically foresees only files transfer.
 The selection of the data of interest is usually performed acting on a few
 parameters (e.g. object name or coordinates). In a few cases, when large
 amounts of data are involved, no (or little) data transfer is allowed but
 the user can
 submit batch jobs that return the results of a particular analysis.
 In other, less common, cases the data are delivered to the user on tapes,
 DVDs, etc.
 In all cases the data acquisition, archiving, delivering, processing and
 the results accessibility are managed separately. Often the information are
 not collected into relational databases tables and when this happens,
 the delay between the date of collection and the archiving is of the order
 of days or even months. The same happens for the data production logging
 and project documentation. Luckily the use, in many cases, of standard
 file formats like FITS\footnote{See FITS Web page: \textsf{fits.gsfc.nasa.gov}}
 can help to track the data origin and processing status.

International projects like those GRID based (see e.g. \textsf{www.grid.org},
\textsf{www.coregrid.net}, \textsf{grid.infn.it}, \textsf{omii-europe.org}, etc.)
and the International Virtual Observatory Alliance (IVOA -- \textsf{www.ivoa.net})
 represent an effort to give a robust and standard
 framework for data archiving, analysis and retrieval
 to physicists and astronomers. However these projects size and ambitions
 cause them to proceed quite slowly and the potential users do not get
 immediate advantages from them.
 Large ground and space based Observatories usually put some effort into
 observations bookkeeping and data accessibility by the users.
 Small and medium projects/experiments instead tend
 to optimize the data management for their internal use only.

Finally we note that the usage of standard data format have allowed
 the development of standard analysis packages, which eventually can be
 easily adapted to meet the requirements of new projects.

\section{Databases in astronomy}
The usage of databases to store data collected by astronomical
instruments/experiments is very common. Still, in the majority of the
cases, they simply contain the information about the collected data
(date, object, wavelength, etc.) or/and a list of objects with their
observed and derived characteristics (catalogues). In a few cases
some level of remote processing is permitted (see e.g. ASDC -- \textsf{www.asdc.asi.it}).
Moreover accessing these information is permitted only via Web browsers
(or http client emulators) or via dedicated programs which typically
also require the data to be on the same machine where the program runs.
Also when very advanced databases systems were implemented, like the one
used by the SDSS
project (\textsf{www.sdss.org}) which allows a direct access (again via http)
with user built SQL queries, a ``direct" communication between the user
program and the database system is not allowed. One needs
for example to submit the query, collect the output into an ASCII file and then
perform all the other desired analyses on his own machine.
 
The Virtual Observatory project is aimed at
removing the obstacles users have in finding and accessing the data,
(cross)processing them and at last retrieve the results,
whatever they are: images, plots, tables,
etc. Still it does not foresee a ``low-level" user interaction.

But why is it so important to make extensive use of databases in Astronomy?
Here is a short list of answers:
\begin{itemize}
\item can track in an ordered form what a project produces and let the rest of
 the world know it;
\item can manage all the information aspects of a project within a single
 framework;
\item don't need to worry about data management but concentrate on the analysis
 and interpretation;
\item make data accessibility uniform for all the Observatories/Projects from
 any computer on the Internet.
\end{itemize}

\section{Our proposed system: MCS}

As mentioned above, the basic idea is that it is easier and more efficient
 to use databases
 for almost all the aspects related to a modern experiment/project in
 Astronomy. Archives with documents, s/w packages, data logs, observing
 schedules, objects catalogues, simulated data, quick-look graphs,
 raw and processed data, \emph{all} can be managed by a modern database
 server
 without caring about computer architecture, programming language,
 access security and even about data sharing, backup and restore.

What do we propose? A system with users management,
multi-threading capabilities, customizable, allowing inter-process messaging
and file transfer, DB input/output from any internet node and using any
programming language.
Database insert/select queries can be performed by mapping the data
into parameters arrays or
structures (generally seen as tables with columns of different types) or
files of various types including FITS, VOTable. Selected data can
be filtered through s/w components producing graphs in vector
(e.g. Postscript) or bitmap format (e.g. GIF), and so on.

Such a system would be also very appropriate to manage experiments in
 real time. Health and data acquisition status, automatic analysis
 results can be monitored from any place on the internet.
The main advantages for a project collaboration would be:
 information easy to find,
 ready to use pre-processed data, shared
 high level processing s/w (automatic or on demand),
 per user backup and restore,
 data access security and easy replication. And again
 the users can have direct access to such a system by using
 custom s/w or Web based user interfaces.
 In other terms,
 the common tasks are performed on the server side whereas clients
 s/w (running on the user computer) can concentrate on specific analysis
 on the retrieved data.

\subsection{The MCS library}
In the last years we have been developing a package which meets
 \emph{all} the above listed requirements: \textsf{MCS}
 (Calderone \& Nicastro, this proceedings).
An MCS based data manager system has the characteristics
of a traditional DB based manager system but with the addition of
several crucial advantages.
It is flexible enough to allow users to easily and quickly
develop tools to manage observation schedules and logs,
real time data archiving and processing.
It has a built-in user's privilege system, SSL encryption and
automatic management of commonly used file formats.
In addition it allows users to easily distribute the data processing
among various machines and keep track of the status via DB log tables.
The MCS library has an interactive shell and it
is interfaced toward (almost) all programming languages; this means that
whereas an MCS server has to be written in C++, any other DB accessing
program can be written in any language.
This permits an easy integration of existing and newly developed
s/w within a collaboration where the participants most likely don't use
one single programming language.

We have also started including user
contributed (MCS based) and external libraries in the various languages
to make even easier to perform DB communication and typical astronomical
analysis/calculations like simple fitting, sky mapping, coordinates
and time conversions, astrometric calculations.
These libraries include well known and tested packages like:
\begin{itemize}
\item Hierarchical Triangular Mesh (HTM -- \textsf{www.sdss.jhu.edu/htm/})
  used for object catalogues indexing;
\item Hierarchical Equal Area isoLatitude Pixelization (HEALPix --
  \textsf{healpix.jpl.nasa.gov}) used to produce sky maps;
\item Naval Observatory Vector Astrometry Subroutines (NOVAS --
  \textsf{aa.usno.navy.mil/software/novas/}) used for
  computing astrometric quantities and transformations.
\end{itemize}
This in one single library which, in his simplest form, can be compiled with
one single dependence: the MySQL (freeware) library. It is also worth noting
that in the future we plan to support DB systems other then MySQL.
Data input/output can be performed in several standard formats like XML,
 FITS and VOTable. The latter immediately makes accessible data and
 products to a Virtual Observatory (VO). Noticeably communication between an
 MCS server and a VO allows to get real time view and access to the
 Observatory products. In other words the Virtual meets the Real.
 We plan to perform ``real" tests in collaboration with Institutions
 involved in the IVOA in the near future.

\section{REM--ROSS}
REM (Rapid Eye Mount, \cite{rem}) is a robotic telescope equipped with
IR (REMIR) and optical (ROSS, \cite{ross})
cameras aimed mainly at catching GRBs afterglows as fast
as possible. It is also used to monitor variable objects and to perform
ToOs observations of other interesting objects.
ROSS can produce direct or dispersed (via an Amici prism) optical images.
 Observation logging and real-time image processing/archiving is performed
 accessing local and remote DBs.
 As soon as the image is (pre)processed, it is available to the owner in
 the database. As usual it is accessible from any internet node.
 A web interface (written in PHP)
 allows a simple and fast access to the log and products (images and spectra).
 Each user has
 his/her own account and can access only proprietary data whereas the
 observation log is freely accessible.
 It is very easy to implement new facilities performing more tasks on images
 or spectra.

Moreover all the REM project documents, papers, pictures, etc. are stored into
DB tables and are accessible through PHP written dynamic web pages.
In addition people have a web accessible ``work area'' repository useful
to exchange any kind of file. The REM observation scheduling and status
information system (see \textsf{ross.iasfbo.inaf.it/$\sim$trem/}),
which was initially implemented to work with ASCII files
rather than with DB tables, will soon start work also in the MCS environment.

\subsection{HTM indexed catalogues}
In order to quickly access IR/optical objects catalogues to discriminate
newly discovered objects in the observed fields, we have ported
into DB tables many of them. The only relevant difference respect to the
original ones is the fact that they are all indexed with the HTM scheme,
which in turn allows a natural DB indexing of the tables. Typically
a query to a one billion objects catalogue like the GSC 2.3, on a
$10\times10$ arcmin area (which is the REM field of view), takes $\sim 20$ ms.
Thanks to MCS, these catalogues can be queried by any (authorized)
internet user directly with his own program, written in any language. 
A standalone program (written in C++) is available to perform simple
\emph{select} queries and get the result in various formats. Moreover a web
interface (written in PHP) allows users to perform interactive queries
with graphical visualization of the selected objects.
All the catalogues are accessible at \textsf{ross.iasfbo.inaf.it}.

\subsection{The ROSS images manager}
\begin{figure}
\centerline{\psfig{file=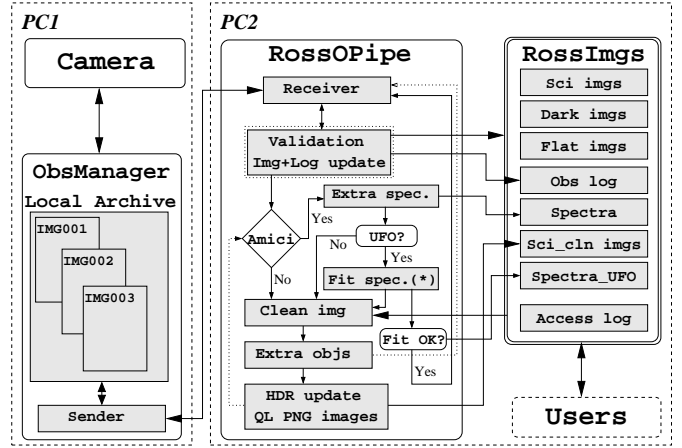,width=8.8cm,clip=} }
\caption{Schematic diagram of the ROSS observing s/w. Access to the MySQL
 custom, all sky catalogues, is performed when searching for UFOs.
\label{rosspipe}}
\end{figure}

The ROSS camera manager is in charge of setting the observation
parameters and performing the images collection as FITS files. Another
s/w component (RossOPipe) manages the objects extraction and matching
with the list
of objects present in the reference (or other) catalogues (e.g. GSC 2.3), this
in order to check for the presence of new (we call them UFOs) objects.
A schematic flow chart is shown in Fig. \ref{rosspipe}.
Finally
all the relevant information are collected in DB tables and made immediately
available to the user which can view them either from a web interface or
via custom programs.

\subsection{The Web based data access}
The observation log, images and spectral data can be browsed, (partially)
processed and retrieved in real time via a PHP written web interface.
Again, thanks to MCS, the same tasks can be performed using any other
language, though interpreted languages like PHP or Python are more
suitable for web pages creation.
The advantages of having a centralized archiver/processing system with an easy
access guarantees:
\begin{itemize}
\item minimum disk space occupancy (in general only post-processing results
 need to be transfered on the user machine);
\item easy backup / restore;
\item easy s/w maintenance.
\end{itemize}
  
\begin{figure}
\centerline{\psfig{file=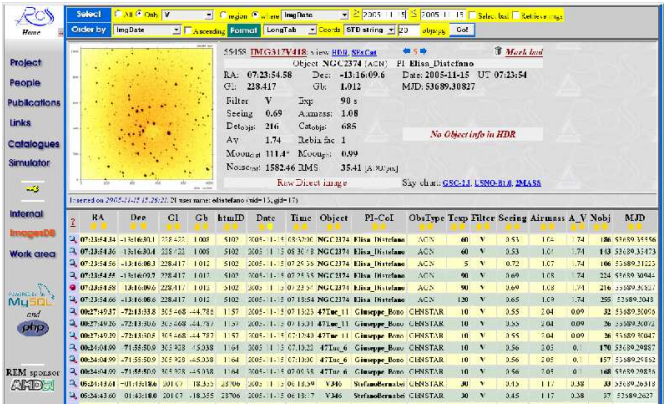,width=8.8cm,clip=} }
\caption{The ROSS images browser uses a simple Web interface.
 Getting information about images and detected objects requires a few clicks.
 Here a V observation of the AGN NGC 2375 is shown.
\label{imgs-browse}}
\end{figure}

\begin{figure}
\centerline{\psfig{file=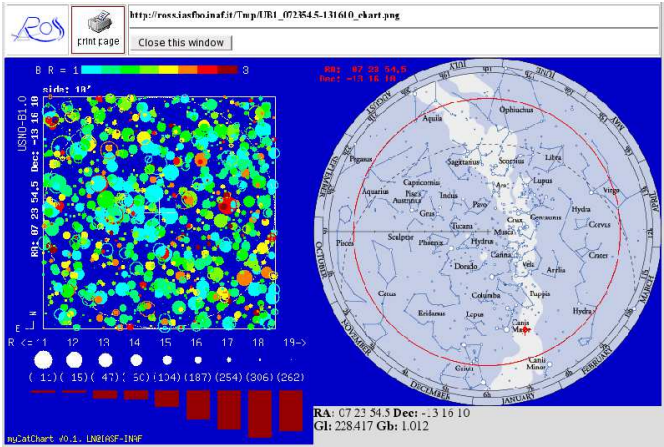,width=8.8cm,clip=} }
\caption{To view the objects present in various (MySQL converted tables)
the Perl written tool \textsf{myCatChart} is used. Here the USNO B1.0 objects
in a $10'\times 10'$ region around NGC 2375 are shown.
It's on the Galactic plane!
\label{sky-chart}}
\end{figure}

\begin{figure}
\centerline{\psfig{file=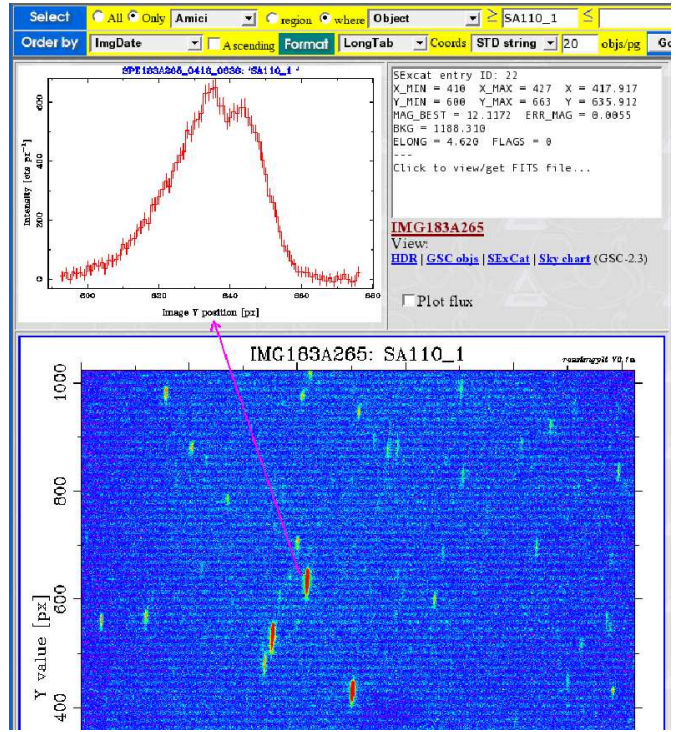,width=8.8cm,clip=} }
\caption{Browsing a spectral image and getting the plots and FITS files cannot
be easier: just click on the image/spectrum.
\label{spe-browse}}
\end{figure}

A record level privilege system allows a selective view of the data in the
database table. Each user can only view and access owned images together
with the calibration files and the observation log. Selection of a sub-sample
of images and browsing/viewing the images on the Web interface is
very simple (see Fig. \ref{imgs-browse}).
Also getting the list of objects in the image with their
photometric and astrometric characteristics requires one click.
Again only one click to view the sky chart of the objects listed in various
all-sky catalogues (see Fig. \ref{sky-chart}).
The automatic spectral data analysis requires only to click on the
spectra to get them plotted and have the corresponding FITS binary tables
in counts or flux units ready to be viewed/downloaded
(see Fig. \ref{spe-browse}). 
Note that \emph{all} these operations are performed in real time.

\section{Conclusions}
We have proposed a new approach to the management of the (nowadays) huge amount
of data/information modern astronomical experiments produce. In particular we
have proposed the usage of a \emph{single} package (\textsf{MCS}, Calderone \&
Nicastro, this volume) which allows users to manage
all the aspects of a project, all built over a relational database system.
Such data would then be more effectively exploitable by the astronomical
community at large for example for multi-wavelength studies and for access
from the various Virtual Observatories.

We welcome any interested group or single researcher willing to contribute
in any aspect of this project.

\vskip 0.4cm
\begin{acknowledgements}
The REM Observatory is supported by INAF.
\end{acknowledgements}
   

        \clearpage

\end{document}